\documentclass[showkeys]{revtex4}
\usepackage{graphicx,amsmath,amssymb,mathrsfs}
\usepackage{multirow}

\renewcommand{\eqref}[1]{(\ref{eq:#1})}

\begin{document}

\title{Superstatistical analysis of sea-level fluctuations}

\author{Pau Rabassa and Christian Beck}

\affiliation{
School of Mathematical Sciences, Queen Mary University of London,
Mile End Road, London E1 4NS, UK}

\begin{abstract}

We perform a statistical analysis of measured time series of sea
levels at various coastal locations in the UK, measured at time
differences of 15 minutes over the past 20 years. When the
astronomical tide and other deterministic components are removed from the
record, a stochastic signal corresponding to the meteorological component
remains, and this is well-described by a superstatistical
model. We do various tests on the measured time series, and compare
the data at 5 different UK locations. Overall the
$\chi^2$-superstatistics is best suitable to describe the data, in
particular when one looks at the dynamics of sea-level {\em differences}
on short time scales.

\end{abstract}


\maketitle

\section{Introduction}

Many complex nonequilibrium phenomena can be understood as the superposition of
different dynamics on different time scales. The superstatistical
approach \cite{beck-cohen} models these complex phenomena as the
superposition of two random variables, one that corresponds to local
equilibrium statistical mechanics and another one that corresponds to a
slowly varying parameter $\beta$, which for example can be the local inverse temperature
in a spatio-temporal inhomogeneous system,
or some local variance parameter associated with slices of a given size in a given time series.
In nonequilibrium statistical mechanics, this
technique
is a powerful tool
to describe a large variety of complex systems
for which there is a spatial or temporal change of conditions on a large scale, larger than
the local relaxation time
\cite{swinney,touchette,chavanis,jizba,
frank,celia,straeten,mark,hanel}. Effectively, when integrating out
the fluctuations of the parameter, this leads to more general types
of statistical mechanics, described by more general entropy functions \cite{tsallis1,hanel,souza}.
Essential for superstatistical models is sufficient time scale separation,
i.e. the local relaxation time of the system must be much shorter than the
typical time scale on which the parameter $\beta$ changes.
Many interesting applications
of the superstatistics concept have been worked out for a variety
of complex systems,
for example the analysis of
train delay statistics \cite{briggs}, 
hydrodynamic turbulence \cite{prl}, cancer survival statistics
\cite{chen} and much more
\cite{daniels,soby,dixit,abul-magd,rapisarda,cosmic}.
Typical distributions of the superstatistical parameter $\beta$ that occur in many of these applications are
the $\chi^2$, inverse $\chi^2$ or lognormal distribution
\cite{swinney}.
Superstatistics based on $\chi^2$ distributions leads to
$q$-statistics \cite{tsallis1,tsallis-book}, whereas other distributions lead to
more complicated versions of generalized statistical mechanics \cite{hanel}.

Of particular interest are superstatistical techniques to analyse and model the complexity inherent in environmental
time series, such as rainfall, wind, surface temperature, and related quantities.
Rapisarda et al. \cite{rapisarda} and Kantz et al. \cite{kantz} investigated
superstatistical aspects of wind velocity fluctuations.
Yalcin et al. \cite{yalcin} did a data analysis of relevant
surface temperature distributions on the earth,
which is important if one wants to understand the effective statistical mechanics
for thermodynamic devices (or local ecosystems)
that are
kept in the open air outside a constant-temperature environment.
Porporato et al. \cite{porporato} looked at
rainfall statistics and developed a model where the rate parameter
of the underlying Poisson process fluctuates in a superstatistical way.

A central point for the applicability of the superstatistics concept is
the existence of suitable time scale separation of the dynamical evolution, or more
generally the existence of a hierarchy of time scales which are well
separated. In the simplest case this just means there
are two different time scales such that the typical variation of $\beta$ takes place
on a much larger time scale than the local relaxation time of the system that is influenced
by the given temperature environment.
There are tests that can check for a given experimental time series if such a time scale separation
is present \cite{straeten}.

In this paper we will deal with a new application of the superstatistics concept that has not
been considered before,
namely the superstatistical analysis of measured sea-level fluctuations at various spatial locations in the UK.
Of particular interest is:
a) which type of superstatistics is relevant for sea levels;  b)
how the results differ from spatial location to location;  c) how well the
time scale separation criterium is satisfied and d) what the relevant time scales are.
Our data set consists of a record of measured sea levels at
various tide gauge locations in the UK
that were measured every 15 minutes over the past 20 years or so.
In Section
\ref{sec:ss-analysis} we review the derivation and application of the superstatistical
approach and the different tools to check its suitability for
a particular data set, in our case sea-level data. In Section \ref{sec:Analysis} we
analyse the data set of observed sea levels at five different
sites in the United Kingdom, and provide answers to questions a)--d).
Our concluding remarks are given in section IV.

\section{Superstatistical analysis
         of time series.}
\label{sec:ss-analysis}

\subsection{The superstatistical model}
\label{sec:ss-model}

Consider a stochastic process that for a short frame of
time is well described by a Gaussian distribution, but on a longer
time scale the parameter value $\beta$ of this Gaussian fluctuates.
Concretely, for a given value of $\beta$ the conditional probability
distribution is given by
 \[
p(v\,|\, \beta) \sim \exp{(-\frac{\beta v^2}{2})}.
\]
Assume that there exists a probability distribution $f(\beta)$ describing
the fluctuations of $\beta$, then the density function
is expected to behave as
\begin{equation}
\label{eq:density-ss}
p(v) = \int_0^\infty f(\beta) p(v\,|\,\beta)d\beta \sim
 \int_0^\infty  f(\beta) e^{-\beta v^2/2} d\beta .
\end{equation}


There are three different superstatistics distributions which are
commonly found in many applications:
\begin{description}
\item[$\chi^2$-superstatistics,] also known as Tsallis statistics. The function
$f(\beta)$ is given by the $\Gamma$-distribution
\[
f(\beta)= \frac{1}{\Gamma\left(\frac{n}{2}\right)}
\left(\frac{n}{2\beta_0} \right)^{n/2} \beta^{n/2 -1} e^{n\beta/2\beta_0},
\]
where $\beta_0$ is the average of $\beta$ and $n$ is the number of degrees of freedom.

\item[Inverse $\chi^2$-superstatistics.] In this case
$f(\beta)$ is given by the inverse $\Gamma$-distribution
\[
f(\beta)= \frac{\beta_0}{\Gamma\left(\frac{n}{2}\right)}
\left(\frac{n\beta_0}{2} \right)^{n/2} \beta^{-n/2 -2} e^{n\beta_0/2\beta},
\]
where $\beta_0$ is again the average of $\beta$ and $n$ is the number of degrees of freedom
of the inverse $\chi^2$ distribution.

\item[Lognormal superstatistics.] In this last case
$f(\beta)$ is described by the lognormal distribution
\[
f(\beta)= \frac{1}{\sqrt{2\pi} s \beta}
\exp\left(\frac{-\left(\ln \frac{\beta}{\mu} \right)^2 }{2s^2} \right),
\]
where $\mu$ and $s$ are suitable parameters.
\end{description}

\subsection{Tools for data analysis}
\label{sec:tools}

Consider $u=\{u_1, \dots, u_n\}$ as a set of $n$ experimental observations
(or data points in general). The total time series is divided into $N$
slices of length $\Delta$, with $N = \lfloor n/\Delta \rfloor$ where
$\lfloor \cdot \rfloor$ denotes the integer part (or floor) function.
The local moment of order $k$ (of the $l$-th slice) is then defined
as:
\[
\langle u^k \rangle_{\Delta,l} = \frac{1}{\Delta} \sum_{j=1+(l-1)\Delta}^{l\Delta} u_j^k,
\text{ with } l=1,2,\dots, N.
\]
In the following, we will assume that the first moment of the total
time series is zero ($\langle u \rangle_{n,1}= 0$).

\subsubsection{Computation of time scales}

A physical process that is well modeled by a superstatistics has two
different time scales. The short time scale
$\tau$ corresponds to the time that it takes for the system to get
to local equilibrium. The long time scale $T$ corresponds to the
time scale that the system remains in local equilibrium before
it fluctuates to another phase state with a different $\beta$.

Given a time series generated by a superstatistical process,
the autocorrelation function should decay exponentially on a short time scale $\tau$ (see \cite{swinney}). In other words
\begin{equation}
\label{eq:decay-cor}
C_{n,\tau}(u) = e^{-1} C_{n,0}(u),
\end{equation}
where $C_{n,t}(u)$ is the autocorrelation function of the time series $u$ defined as
\[
C_{n,t} (u) = \frac{1}{n-t} \sum_{i=1}^{n-t} u_i u_{i+t}.
\]

Assume that we have our data divided into $N$ slices of length $\Delta$
as before. We can consider the local averaged kurtosis associated
to that length
\begin{equation}
\label{eq:kappa-1}
\kappa_{\Delta} : = \frac{1}{N} \sum_{l=1}^{N} \kappa_{\Delta,l}, \text{ with }
\kappa_{\Delta,l} = \frac{\langle u^4 \rangle_{\Delta,l}}{\langle u^2 \rangle_{\Delta,l}}.
\end{equation}

The long time scale $T$ corresponds to the optimal time scale
for which the process is best described as a superposition of
Gaussian random variables. The choice proposed in \cite{swinney}
is to take $T$ such that
\begin{equation}
\label{eq:kappa-equals-3}
\kappa_{\Delta} = 3.
\end{equation}
This choice has been found to work well for a variety of experimental
data \cite{swinney,straeten}.

\subsubsection{An additional constraint}

Besides the existence of two separated timescales, it is also necessary to check
that the Gaussian hypothesis used to derive the model is satisfied.
Assume that we have found the value of $\Delta$ for which $\kappa_\Delta = 3$; we
denote it by $T$. The original time series is divided into $N = \lfloor n/T \rfloor$
slices of length $T$. Let $\langle u^2 \rangle_{T,l}$ and  $\langle u^4 \rangle_{T, l}$
denote the second and the fourth moments of these time slices.

Consider the value $\kappa_{\Delta}$ defined by  \eqref{kappa-1} for
$\Delta = NT$. For large data series, $\kappa_{NT}$ is a
good approximation of $\kappa_{n,1}$ (the kurtosis of the complete data set)
because $N = \lfloor n/T \rfloor \approx n/T$.
Moreover,  $\kappa_{NT}$ can be expressed as follows:
\begin{equation}
\label{eq:kappa-2}
\kappa_{NT} = \left( \frac{1}{N} \sum_{j=1}^N \langle u^2 \rangle_{T,j}\right)^{-2}
 \frac{1}{N} \sum_{j=1}^N \langle u^4 \rangle_{T,j}
\end{equation}

In the case that the data in each of these time slices are distributed
exactly like a Gaussian random variable
(with zero mean and variance $1/\beta_{\Delta,l}$), then
the first four local moments are
\[
\displaystyle
\langle u \rangle_{T,j} = 0,  \quad
\langle u^2 \rangle_{T,j} = \frac{1}{\beta_{\Delta,j}}, \quad
\langle u^3 \rangle_{T,j} = 0, \quad
\langle u^4 \rangle_{T,j} = \frac{3}{\beta_{\Delta,j}^2}.
\]

In the case where the data in each of these time slices is approximately
distributed like a Gaussian random variable, we expect the
moments to be close to the ones above. Let $\theta_{T,l}$
be the deviation of the fourth moment $\langle u^4 \rangle_{T,l}$
from $3 \langle u^2 \rangle_{T,l}^2$ for the $l$-th time slice of length T
\[
\theta_{T,l} = \langle u^4 \rangle_{T,l} - 3 \langle u^2 \rangle_{T,l}^2.
\]

The approximate kurtosis \eqref{kappa-2} can be rewritten as:
\[
\kappa_{NT} = \left( \frac{1}{N} \sum_{j=1}^N \langle u^2 \rangle_{T,j}\right)^{-2}
 \frac{1}{N} \sum_{j=1}^N \left[ 3 \langle u^2 \rangle_{T,j}^2  + \theta_{T,j}\right].
\]

When the Gaussian approximation is reasonable, the contribution of the terms $\theta_{T,j}$
to $\kappa_{NT}$ is small as compared to the terms $3 \langle u^2 \rangle_{T,j}^2$. Consider the
parameter $\epsilon$ given as
\begin{equation}
\label{eq:epsilon}
\epsilon = \frac{1}{3}
\frac{ \sum_{j=1}^N \theta_{T,j} } { \sum_{j=1}^N  \langle u^2 \rangle_{T,j}^2}.
\end{equation}
This parameter measures the contribution of the deviations from the Gaussian
approximation in these time slices to the value of $\kappa_{n,1}$.
Concretely,  we expect the local Gaussian approximation to
hold for a given time series when $|\epsilon| $ is small. For more details see
\cite{straeten}.


\section{Analysis of the sea level data}
\label{sec:Analysis}

\subsection{Nature of the sea level data}

\begin{figure}[t]
\begin{center}
\includegraphics[width=0.5\textwidth]{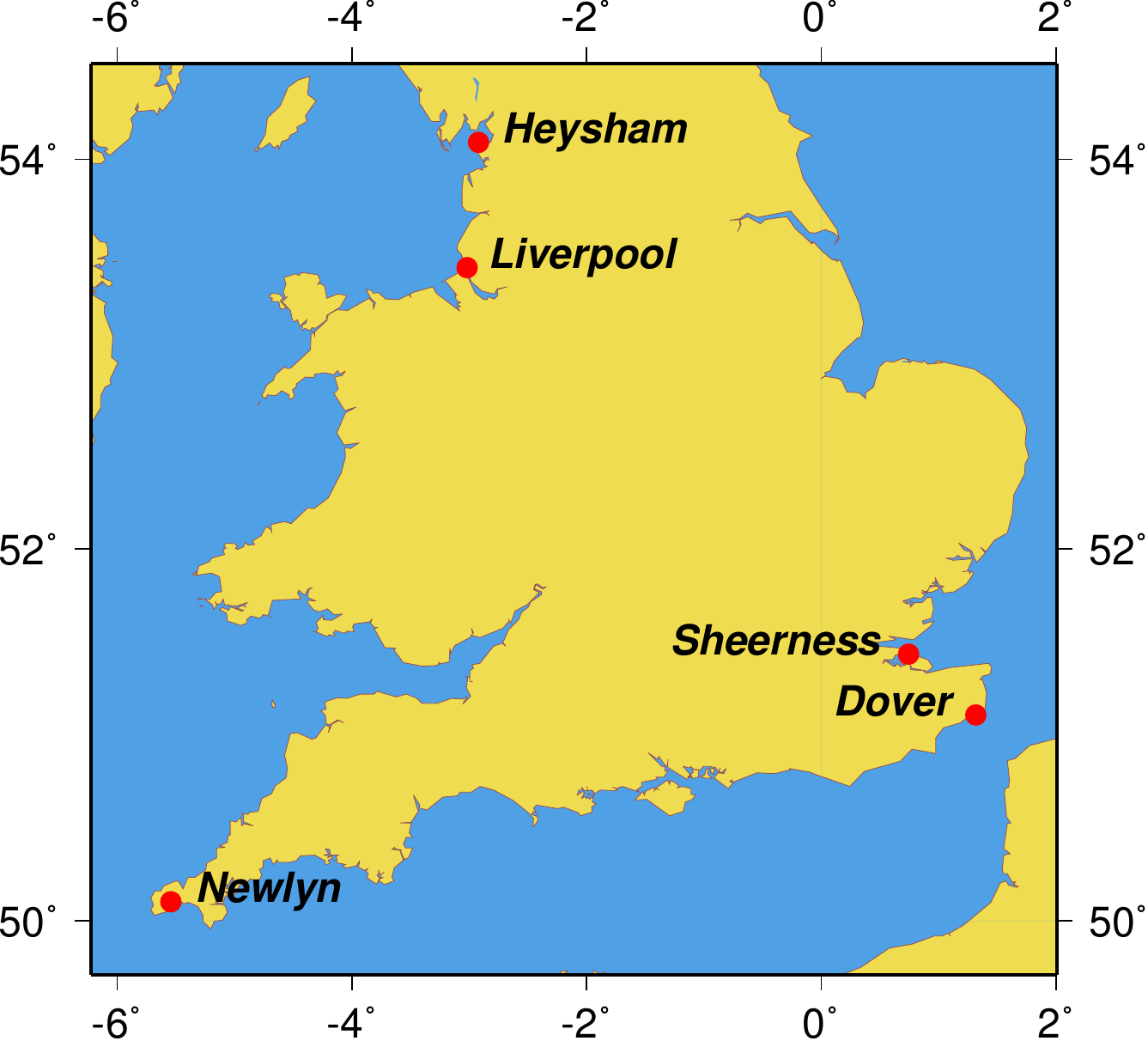}
\end{center}
\caption{Geographic locations of sea level gauge stations considered.}
\label{fig:0}
\end{figure}

\begin{figure}[t]
\begin{center}
\includegraphics[width=0.76\textwidth]{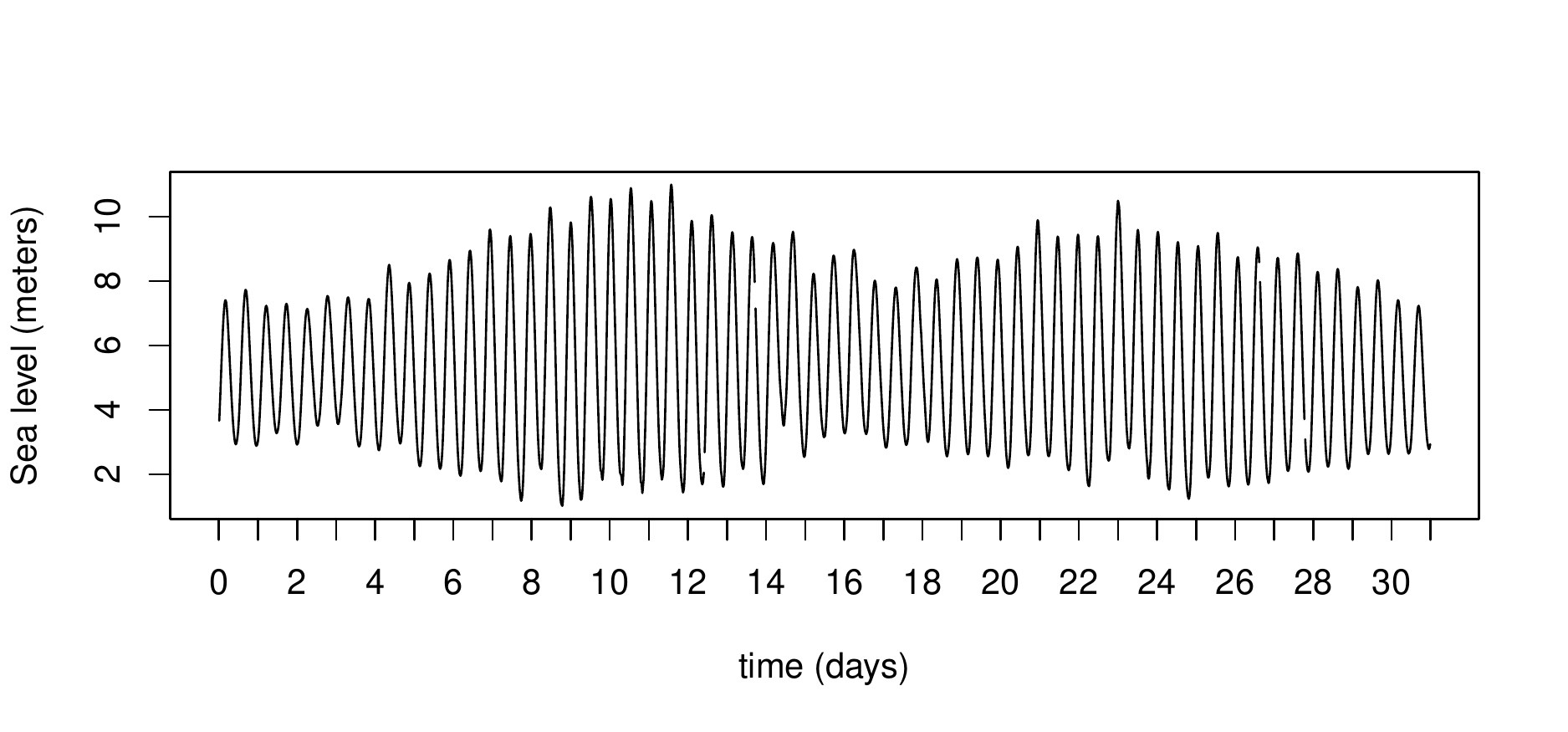}
\includegraphics[width=0.76\textwidth]{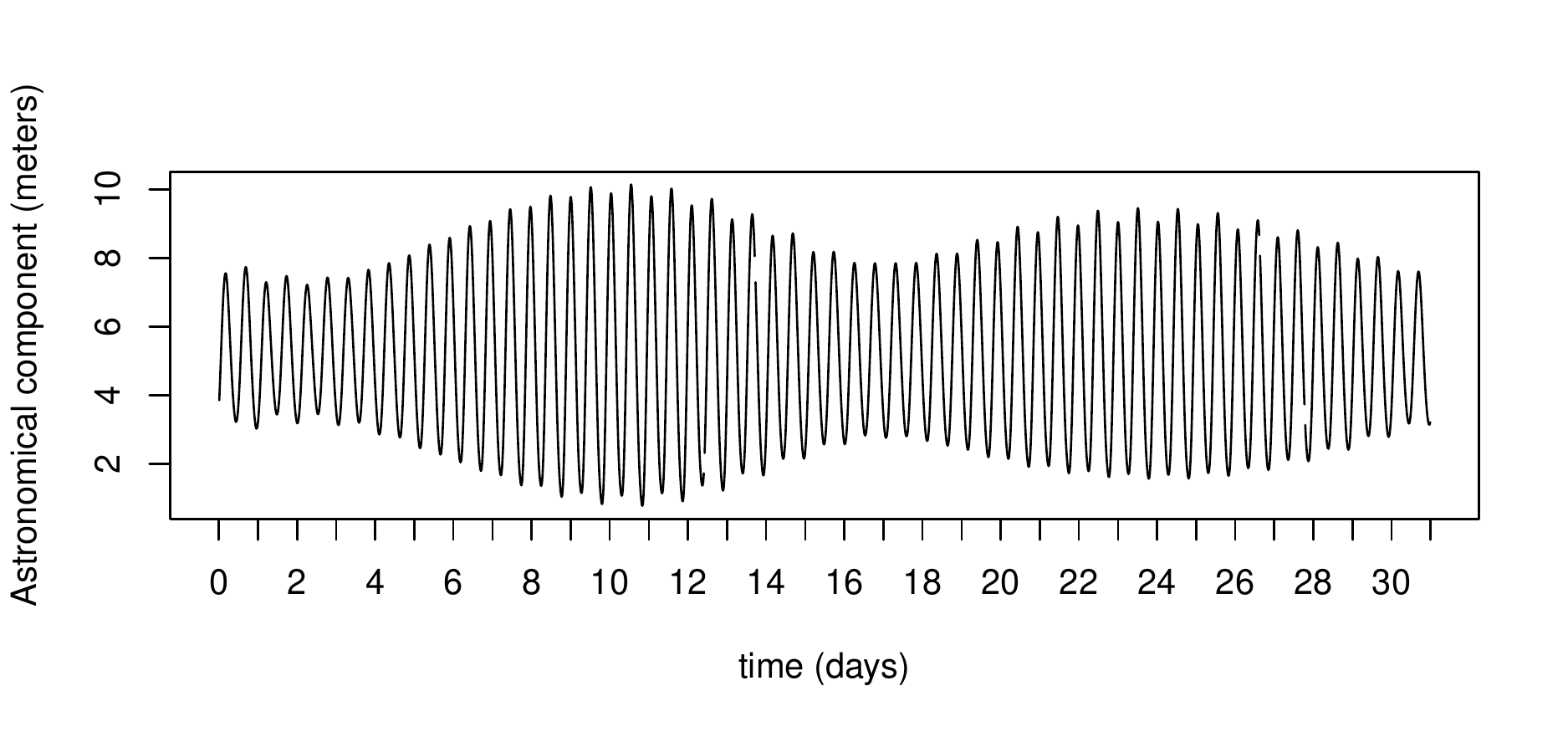}
\includegraphics[width=0.76\textwidth]{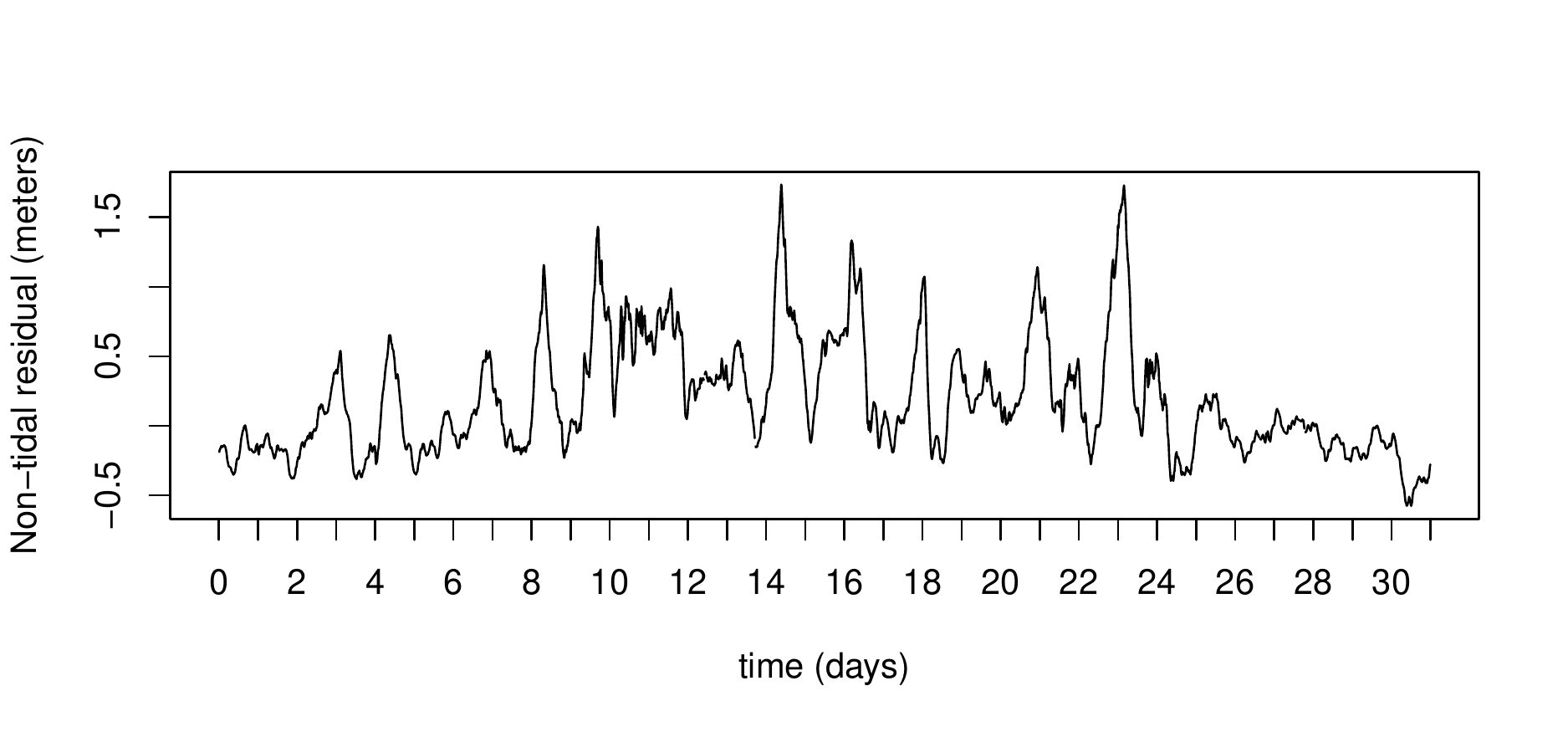}
\end{center}
\caption{Sea-Level time series at Heysham in January 1993.
Top: observed sea level $Z(t)$. Middle: astronomical tide $X(t)$.
Bottom: Non-tidal residual $Y(t)$.}
\label{fig:1}
\end{figure}

\begin{figure}[t]
\begin{center}
\includegraphics[width=0.76\textwidth]{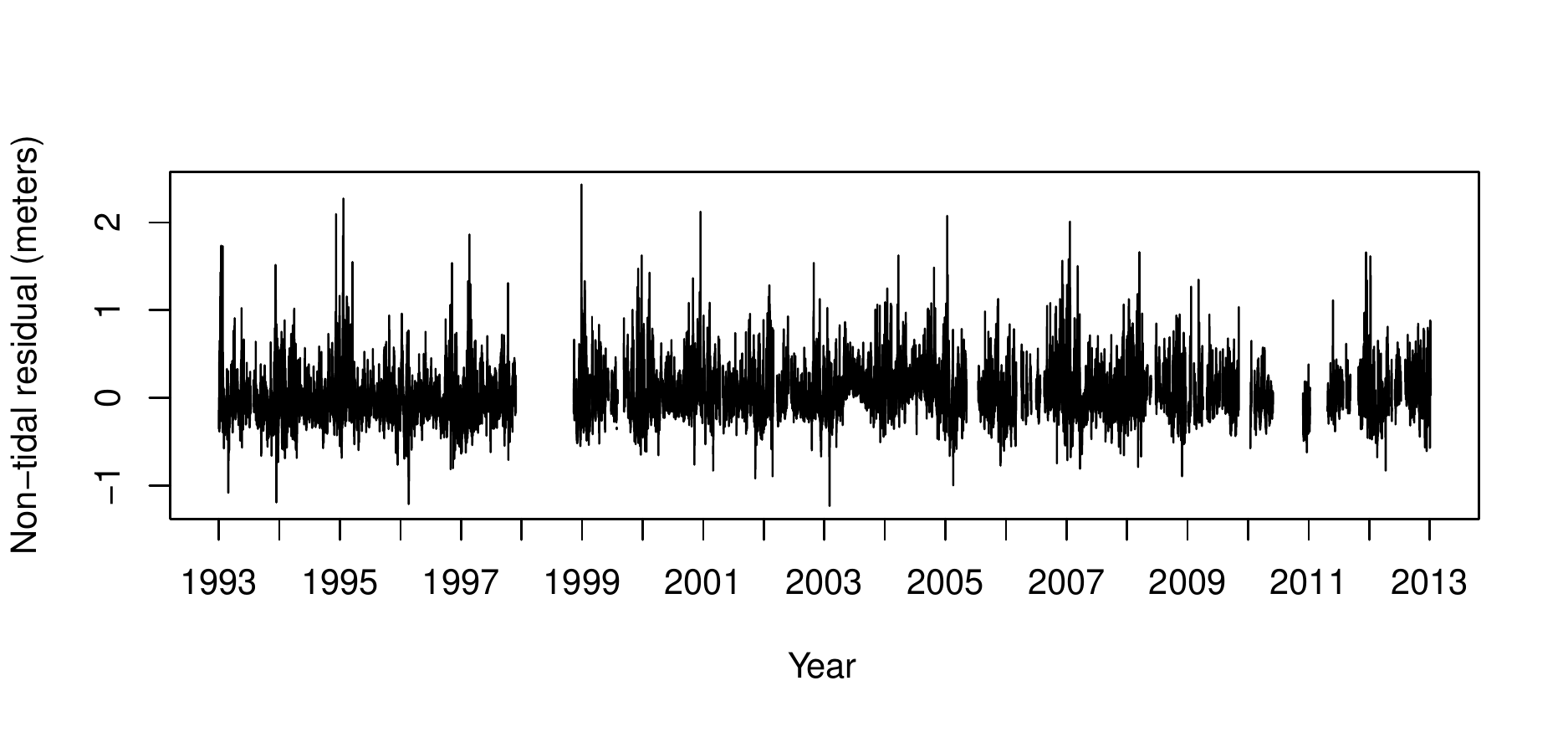}
\end{center}
\caption{Available non-tidal residual $Y(t_{i})$ at Heysham
from 1 January 1993 to 31 December 2012.}
\label{fig:2}
\end{figure}

\begin{figure}[t]
\begin{center}
\includegraphics[width=0.76\textwidth]{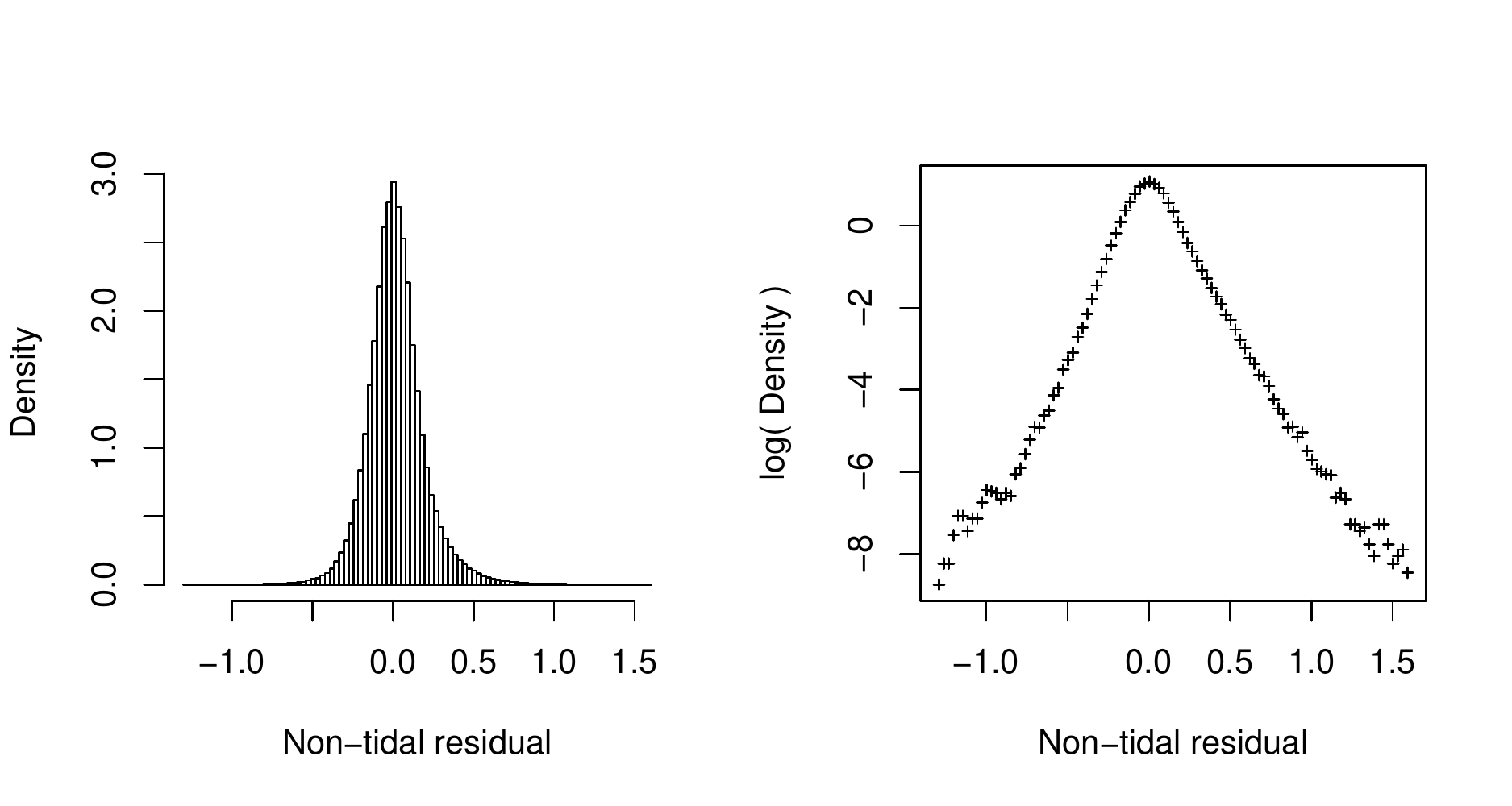}
\end{center}
\caption{Histogram of non-tidal residuals at Dover
(in a logarithmic scale at the left).}
\label{fig:3}
\end{figure}

\begin{figure}[t]
\begin{center}
\includegraphics[width=0.76\textwidth]{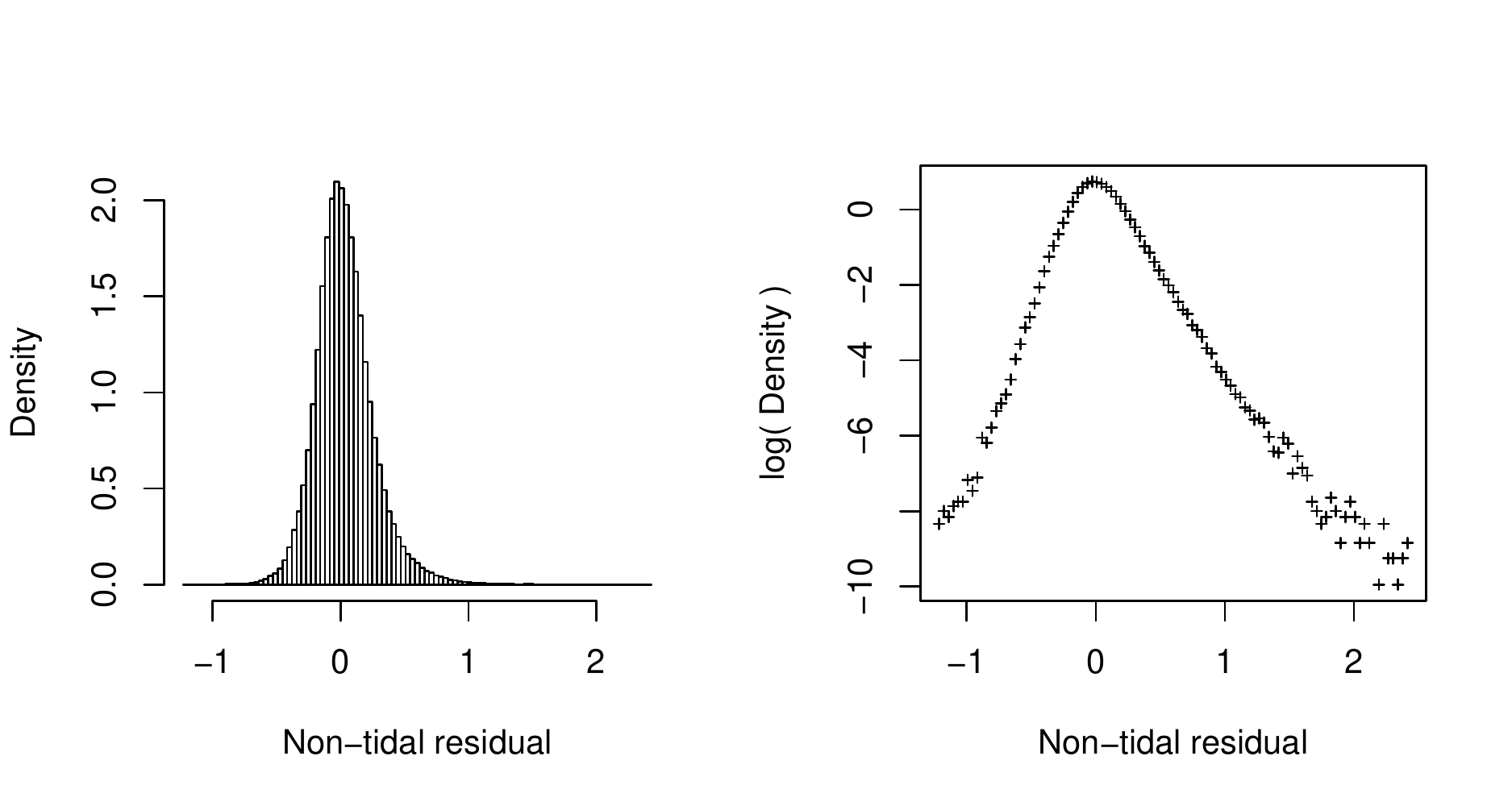}
\end{center}
\caption{Histogram of non-tidal residuals at Heysham
(in a logarithmic scale at the left).}
\label{fig:4}
\end{figure}

\begin{table}[t]
\begin{center}
\begin{tabular}{| l ||  c | c | c |}
\hline
Gauge location       & Recorded  & Missing  & Completeness \\
\hline
Dover      & $640225$ & $61055$  &  $91.29\% $ \\
Heysham    & $562896$ & $138384$ &  $80.27\% $  \\
Liverpool & $616066$ & $85214$  &  $87.85\% $ \\
Newlyn     & $654818$ & $46462$  &  $93.37\% $ \\
Sheerness  & $526026$ & $175254$ &  $75.01\% $ \\
\hline
\end{tabular}
\end{center}
\caption{Complementary information to the analysed data. We have the
number of satisfactorily recorded observations, the number of missing observations
and the \% of completeness for each site.}
\label{tb:0}
\end{table}

\begin{table}[t]
\begin{center}
\begin{tabular}{| l |  r  r  r  r  r |}
\hline
       & Dover    & Heysham & Liverpool & Newlyn & Sheerness \\
\hline
Dover      & $1.000$  & $0.061$  &  $0.112$ & $0.144$ & $0.745$ \\
Heysham    & & $1.000$  &  $0.885$ & $0.412$ & $-0.257$ \\
Liverpool &  & &  $1.000$ & $0.379$ & $-0.242$ \\
Newlyn     &  & & & $1.000$ & $0.016$ \\
Sheerness  &  & & & & $1.000$ \\
\hline
\end{tabular}
\end{center}
\caption{Correlation matrix for non-tidal residuals at five
different tide gauges in the UK.}
\label{tb:1}
\end{table}

In this paper we analyse observed sea level data available from the
British Oceanographic Data Centre \cite{BODC}. Our study
focusses on recorded sea levels from the UK Tide Gauge
Network at five particular locations: Dover, Heysham, Liverpool, Newlyn
and Sheerness. These locations have been selected for our analysis
as a representative subset out of a network of $45$
different sites. The map in Figure \ref{fig:0} shows the geographical
locations of these five gauge stations.

The data at each location is composed of one observation every
quarter of hour from the 1st of January 1993 to the 31st of December 2012. The
data is processed and quality-controlled by the British Oceanographic Data Center.
Each satisfactory recorded sea level value is also provided with a corresponding
value of the non-tidal residual (see below for details).  Table
\ref{tb:0} displays some background information on the number of 
data recorded in a satisfactory way at each site for the time period considered.

The observed sea level $Z(t)$ (at a given time $t$) is typically decomposed
into three different components \cite{tawn}
\[
Z(t) = M(t) + X(t) + Y(t),
\]
which are the following:

\begin{description}
\item[Annual mean sea level] ($M(t)$). On a local scale its
temporal changes are the product of a couple of
(inherently coupled) processes such as volume and
mass changes due to (i) exchange of water with the atmosphere
and the continents and (ii) mass redistribution due to changing
winds and sea level pressure \cite{merrifield,calafat,dangendorf1,dangendorf2},
and vertical land motion \cite{peltier, woppelmann}.
These variations vary locally and can be as large as $\sim \pm 80$cm on
a inter-monthly or $\sim \pm 50$cm on an inter-annual time scale.
Additionally there is a secular long-term trend of the mean sea level
from $1$ to $3$mm/year \cite{cabanes, haigh}.
Given a set of observations at a particular site, it can be determined by
simple regression and removed from the data.

\item[The astronomical tide] ($X(t)$). This is the non-stochastic part of
the sea-level change that is caused by the changing forces of moving
planetary objects. This component is modelled as the sum of $m$ different
tidal constituents \cite{pawlowicz}:
\[
X(t) = \sum_{i=1}^{m} H_i \cos(\omega_i t + \phi_i).
\]
The angular frequencies $\omega_i$  and the number of
constituents $m$ are fixed, whereas the amplitudes $H_i$ and
the phase lags $\phi_i$ can be computed from the data on every site.

\item[Non-tidal residual]($Y(t)$). This component primarily contains the
meteorological contribution to sea level often termed the surge,
which has stochastic behaviour. This component can be also
influenced by tide-surge interaction, harmonic prediction error, gauge
timing error, tide-river flow interaction, etc.; see \cite{batstone} and
references therein.

\end{description}

In Table \ref{tb:1} we have included the cross-correlation matrix
associated with observed non-tidal residual levels at the different
locations. We can observe that there is some correlation between the locations
that are geographically close to each other (see Figure~\ref{fig:0}),
Dover correlated with Sheerness and Heysham correlated with Liverpool, whereas there
is only little correlation between distant sites.

Figure \ref{fig:1} displays the observations at Heysham in January 1993.
The figure also shows the deterministic astronomical tide (middle panel)
and the non-tidal residual (bottom panel) for one month. The non-tidal
residual for the whole 20-year period is shown in Figure \ref{fig:2}.

Our analysis focusses on determining if the observed non-tidal residual $Y(t)$
is well fitted by a superstatistical process.
We can observe in Table \ref{tb:0} and Figures \ref{fig:1} and
\ref{fig:2} that the dataset is far from complete, there are both
some single observations missing and long periods of time without data.
For practical purposes, we will consider
only the available data and ignore all the missing values,
which are due to mechanical problems with the tide gauge. Therefore our
data set will comprise the values for which $Y(t_i)$ is available.

Figures \ref{fig:3} and \ref{fig:4} show the histograms of recorded
non-tidal residuals at Dover and Heysham, respectively.
The distributions are non-Gaussian and have fat tails.
Some skewness can also be observed in both cases (see also
Table \ref{tb:2}). This is mainly due to the presence of the
coast, where the water is piled up, which is not the
case if the winds push the water away from the coast.
This makes positive surges more extreme than negatives ones,
producing a skewed distribution. For more details on the processes
contributing to the surge generation in the region see
\cite{calafat, dangendorf2,horsburgh} and references therein.

\subsection{Analysis of the non-tidal residual}
\label{sec:analysis-surge}

\begin{figure}[t]
\begin{center}
\includegraphics[width=0.38\textwidth]{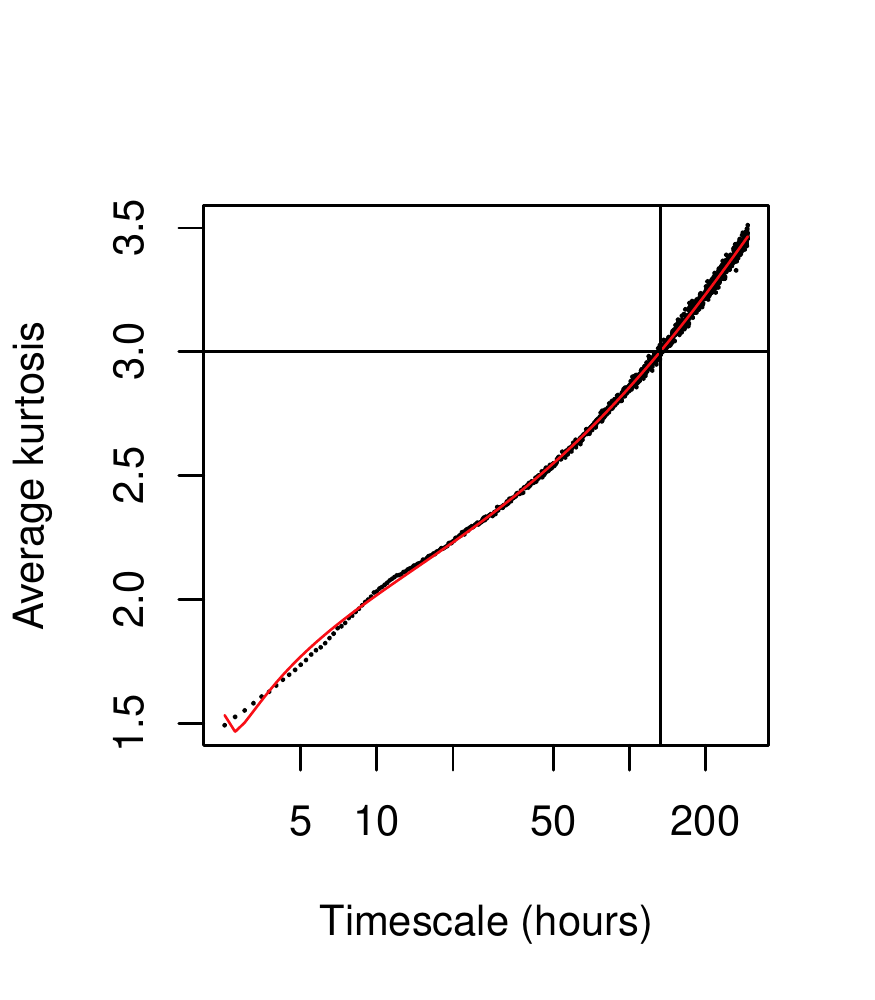}
\includegraphics[width=0.38\textwidth]{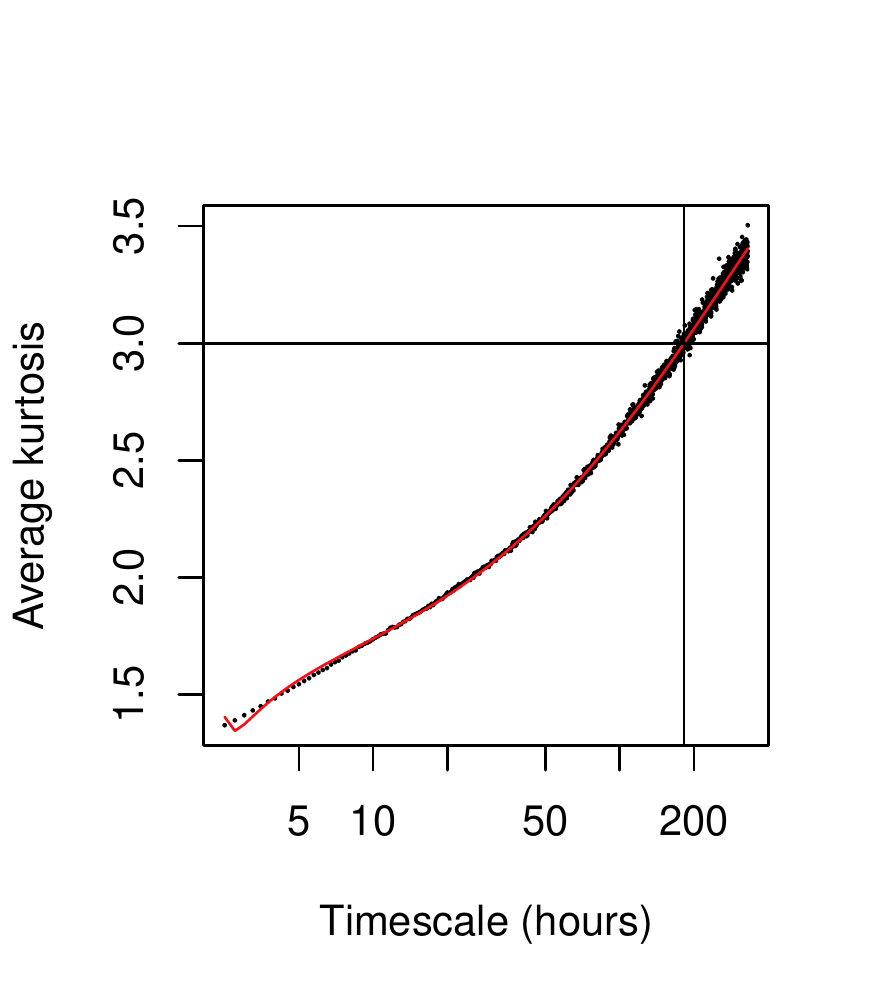}
\end{center}
\caption{Computation of optimal time scale $T$ for observed non-tidal residuals at
Dover (left) and Heysham (right). Every dot corresponds to a value of the
average kurtosis $\kappa_\Delta$ given by \eqref{kappa-1}. The red line is a
polynomial function of degree $5$ that best fits the data points. The optimal
time scale is obtained as the intersection of this line with $\kappa=3$.}
\label{fig:5}
\end{figure}

\begin{figure}[t]
\begin{center}
\includegraphics[width=0.76\textwidth]{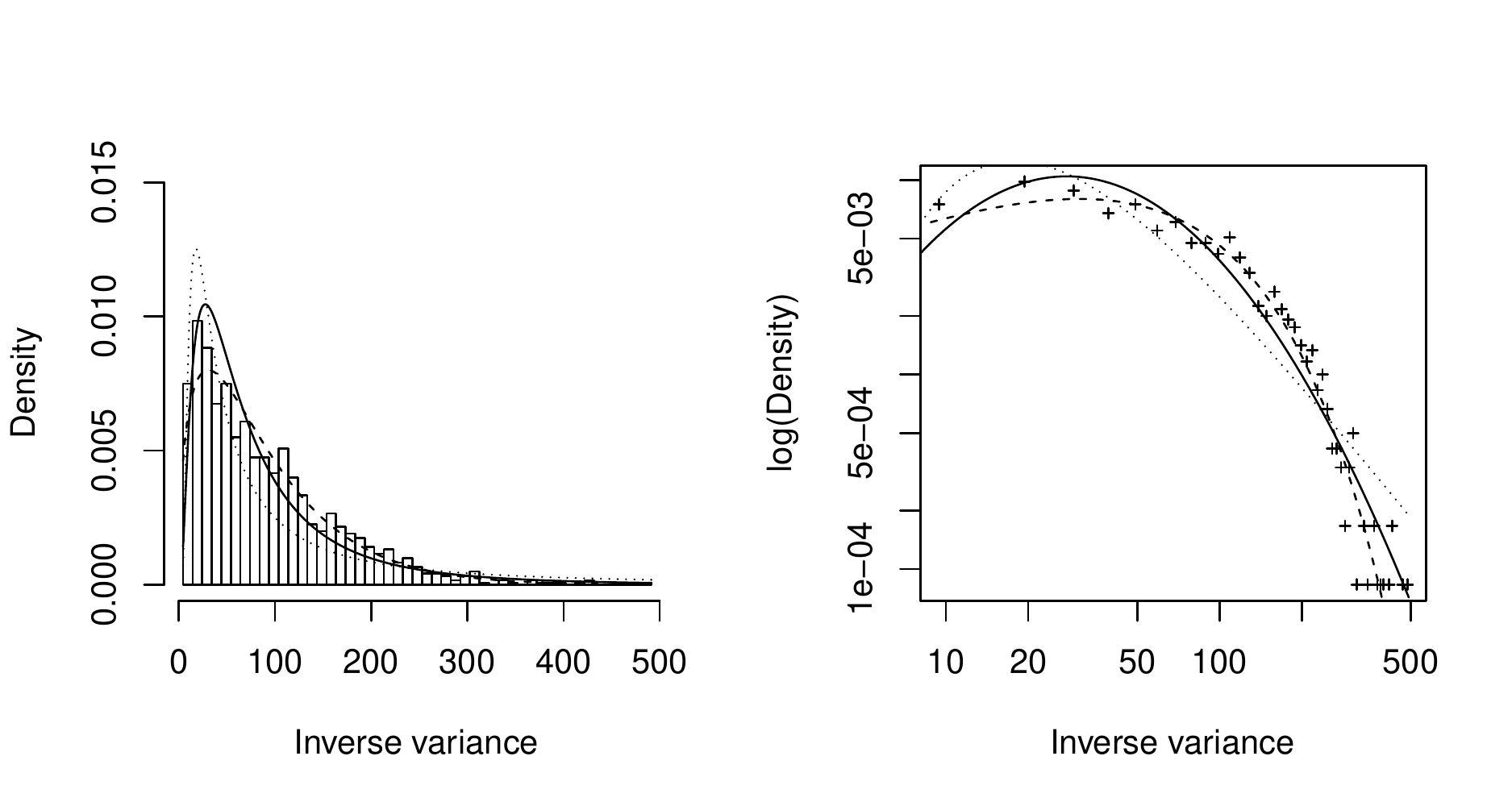}
\end{center}
\caption{Probability density $f(\beta)$ extracted form the non-tidal residuals
at Dover and compared to log-normal (solid line), $\chi^2$ (dashed line) and
inverse-$\chi^2$ (dotted line). Left:linear-linear scale, right:log-log scale.
All distributions have the same mean and variance as the experimental data.}
\label{fig:6}
\end{figure}

\begin{figure}[t]
\begin{center}
\includegraphics[width=0.76\textwidth]{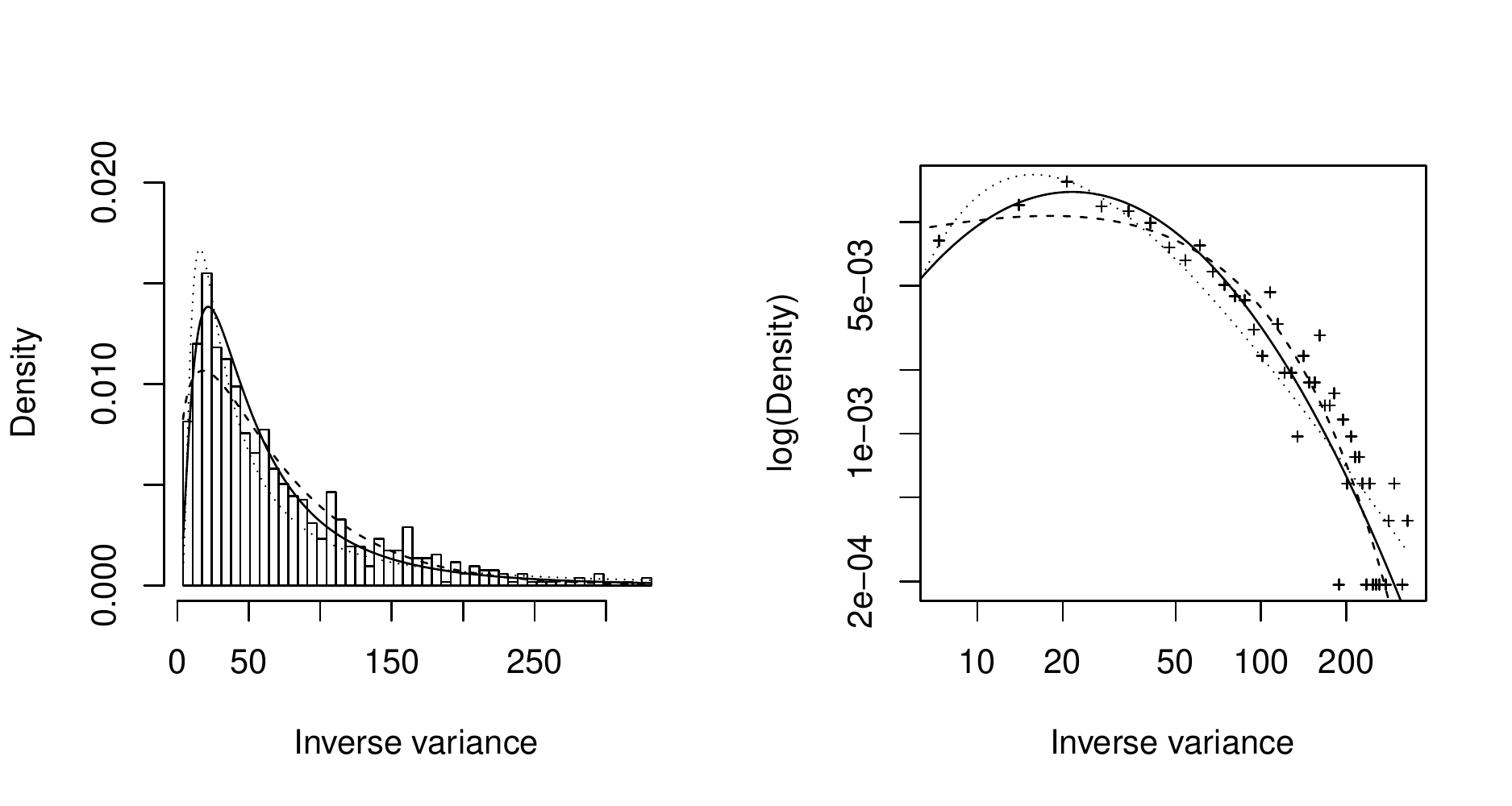}
\end{center}
\caption{As Figure \ref{fig:6} but for non-tidal residuals at Heysham.}
\label{fig:7}
\end{figure}

\begin{table}[t]
\begin{center}
 \begin{tabular}{| l | r | r | r | r | r | r | c | c | c| }
 \hline
 \multirow{2}{*}{Location} & \multirow{2}{*}{ $\tau$} & \multirow{2}{*}{$T$ } &
\multirow{2}{*}{$\tau/T$}& \multirow{2}{*}{$|\epsilon|$} & \multirow{2}{*}{ Skewness} &
\multirow{2}{*}{ Kurtosis}  & \multicolumn{3}{ c|}{ K-S Distance} \\
 \cline{8-10}
 & & & & & & & log-norm. & $\chi^2$ & Inv-$\chi^2$ \\
 \hline
 Dover     &  $13.5$ &  $132.5$ & $0.10$ & $0.08$
& $0.58$ & $6.68$ & $0.062$ & $0.032$ & $0.169$ \\
 Heysham   & $53.25$ &  $182.5$ & $0.29$ & $0.13$
& $0.86$ & $6.32$ & $0.035$ & $0.041$ & $0.117$ \\
 Liverpool & $53$ &  $169.75$ & $0.31$ & $0.16$
& $0.96$ & $6.77$ & $0.057$ & $0.030$ & $0.138$ \\
 Newlyn    & $104.25$ & $998$ & $0.10$ & $0.04$
& $0.54$ & $3.91$ & $0.039$ & $0.048$ & $0.066$  \\
 Sheerness &  $9$ &  $83.75$ & $0.11$ & $0.13$
& $0.24$ & $9.42$ & $0.057$ & $0.033$ & $0.236$  \\
 \hline
\end{tabular}
\end{center}
\caption{Relevant statistical parameters for applicability of the superstatistical
model to observed non-tidal residuals at different locations.
The short and the long time scales ($\tau$ and $T$) are expressed in hours.
Parameters $\tau$, $T$ and $\epsilon$ are described in Section \ref{sec:tools}.}
\label{tb:2}
\end{table}

In order to determine if a data set can be well modeled by a superstatistics based
on locally Gaussian distributions with fluctuating variance,
one should check that certain conditions are fulfilled. The first
one is that the data has suitable high order moments. Skewness should
be close to $0$ and kurtosis should be greater that $3$. The second
conditions is to check that there exists a proper time-scale separation.
The short time scale $\tau$ corresponds to the time that it takes for the system to get
to local equilibrium. It can be computed from the data using eq.~\eqref{decay-cor}.
The long time scale $T$ corresponds to the time scale that the system remains in
local equilibrium before it fluctuates to another state with different $\beta$. This time scale can be
computed as the particular value of $\Delta$ such that eq.~\eqref{kappa-equals-3} holds (see
Figure \ref{fig:5}). The ratio between
the short time scale $\tau$ and the long time scale $T$ should be small enough in
order to have a proper time scale separation. The third condition to check is
that the local Gaussian approximation is a reasonable hypothesis. This corresponds to
checking that the parameter $\epsilon$ given by eq.~\eqref{epsilon} is close to zero.

Once the optimal time scale $T$ has been determined one can extract the
empirical distribution function $f(\beta)$ that defines the
superstatistical model \eqref{density-ss}. For this empirical distribution
one can determine which of the three universality classes described in
Section \ref{sec:ss-model} fits the data best.
One possibility is to do this just by visual inspection
of the fitted distributions \cite{swinney,straeten}, but that
has the inconvenience of being tedious, especially when applied to multiple
datasets or when the parameter space is higher-dimensional. Moreover the visual analysis can be subjective and sometimes
it can be difficult to tell which distribution fits best (see Figures \ref{fig:6}
and \ref{fig:7}). As a more systematic method we here consider the Kolmogorov-Smirnov
(K-S) distance as an objective criterium to determine the best fit. The K-S distance
measures the difference between the empirical distribution and the proposed one
\cite{eadie}.
Given $u=\{u_1, \dots, u_n\}$, a set of observations, its empirical distribution
is defined as $F_n(v) = \frac{1}{n} \sum_{i=1}^{n} I_{u_i\leq v}$, where
$I_{u_i\leq v}$ is the indicator function, equal to $1$ if $u_i \leq v$
and equal to $0$ otherwise. Given a cumulative distribution function $F$,
the K-S distance associated to the dataset $u$ is defined as
\[
D_n = \sup_{v} \left| F_n(v) - F(v) \right|.
\]

In Table \ref{tb:2} we show the results of our analysis, i.e.\  the time scale parameters
$\tau$ and $T$, the parameter $\epsilon$ defined in eq.~(\ref{eq:epsilon}), the kurtosis and
skewness of the histogram of the entire signal, as well as the K-S distance of our superstatistical fits,
for the five different sites considered. The time scale separation is not very
strong, but it is present. The values of $|\epsilon|$ are in general small for all
sites which indicates that the local Gaussian approximation is reasonable. The
kurtosis value is always greater than $3$ for the entire time series. The parameter that disagrees the most
with the hypothesis of simple Gaussian local behavior is the skewness of the data.
The data is skewed due to the non-symmetric nature of the storm surges,
whereas the superstatistical model in its simplest form is deduced as
the superposition of normal random variables
which are not skewed. Of course one can consider locally
non-symmetric correction terms to the Gaussian, as done e.g.
in \cite{hydro}. However, this introduces an additional fitting parameter to recover the skewness.
We avoid this problem in the next section
by considering the dynamics of temporal changes of the non-tidal
residual instead of the residual itself.

Finally we observe that the K-S distance does not shows uniquely which distribution
fits the data best. For all locations the empirical distribution $f(\beta)$
is reasonably well fitted by either a lognormal or a $\chi^2$ distribution, but there is not a
clear favourite (see also Figures \ref{fig:6} and \ref{fig:7}).

\subsection{Analysis of non-tidal residual differences}

\begin{figure}[t]
\begin{center}
\includegraphics[width=0.76\textwidth]{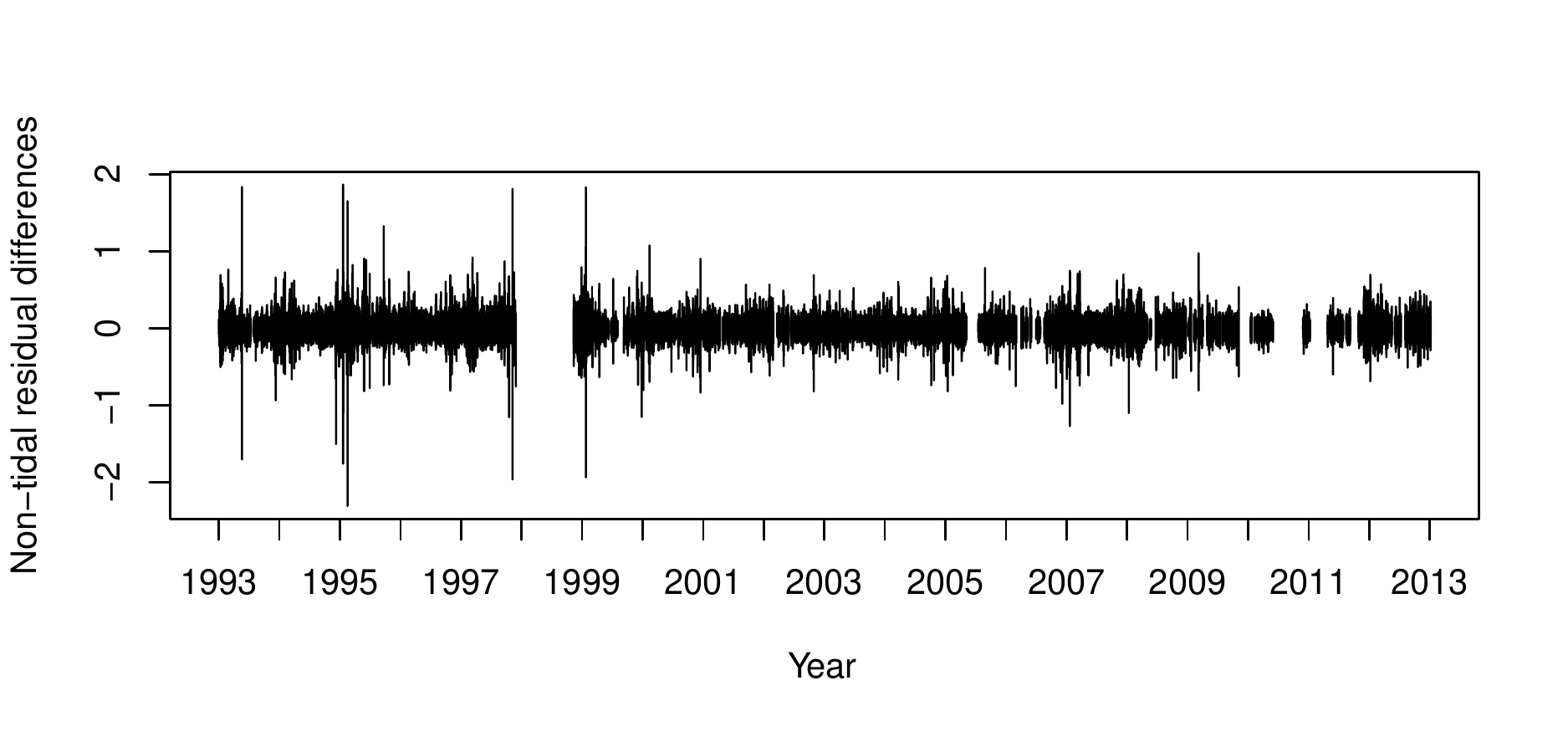}
\end{center}
\caption{Available non-tidal residual differences $Y(t_{i+1})-Y(t_{i})$ at Heysham
from 1 January 1993 to 31 December 2012.}
\label{fig:8}
\end{figure}

\begin{figure}[t]
\begin{center}
\includegraphics[width=0.76\textwidth]{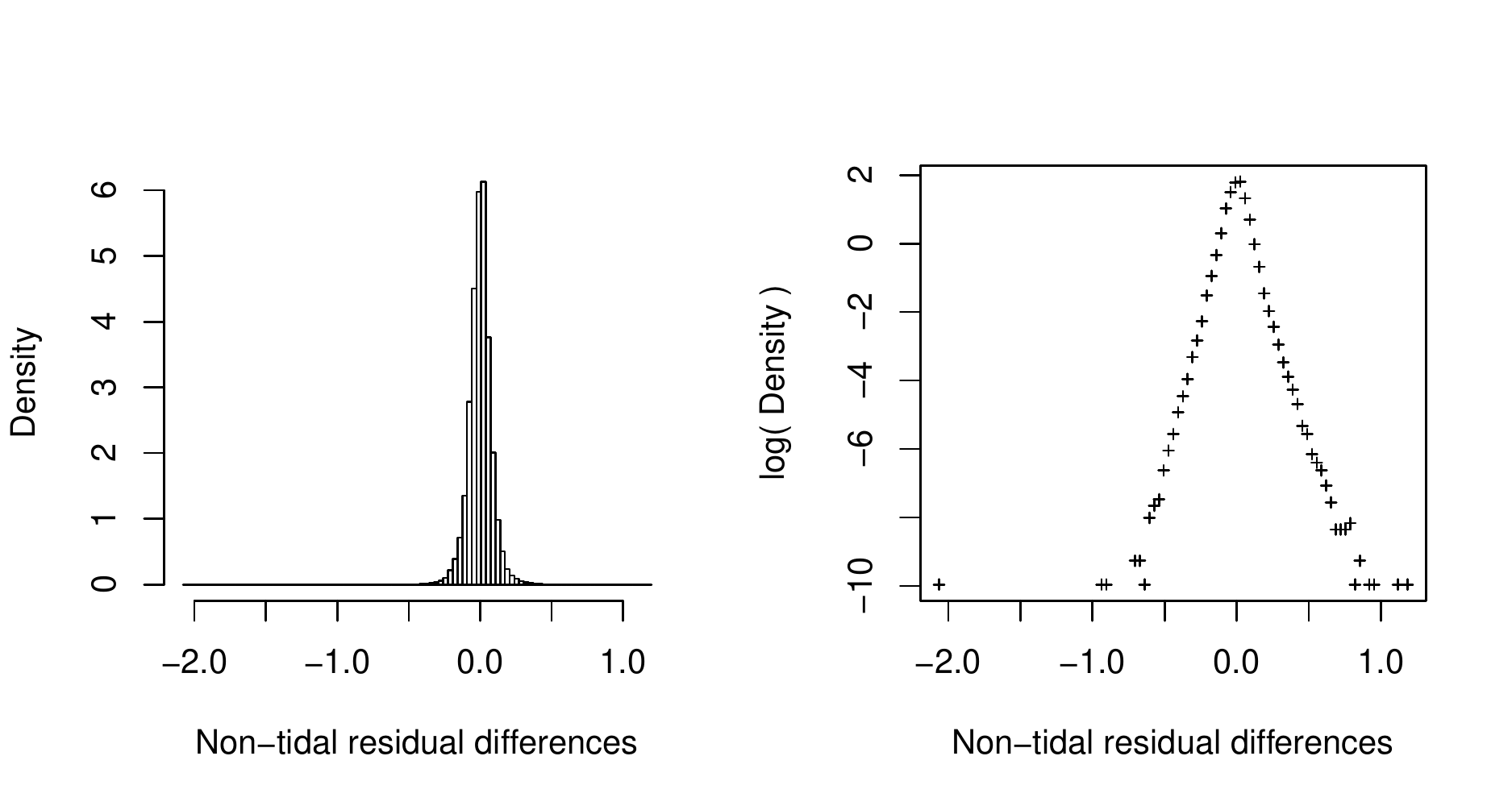}
\end{center}
\caption{Histogram non-tidal residual differences at Dover
(in a logarithmic scale at the left).}
\label{fig:9}
\end{figure}

\begin{figure}[t]
\begin{center}
\includegraphics[width=0.76\textwidth]{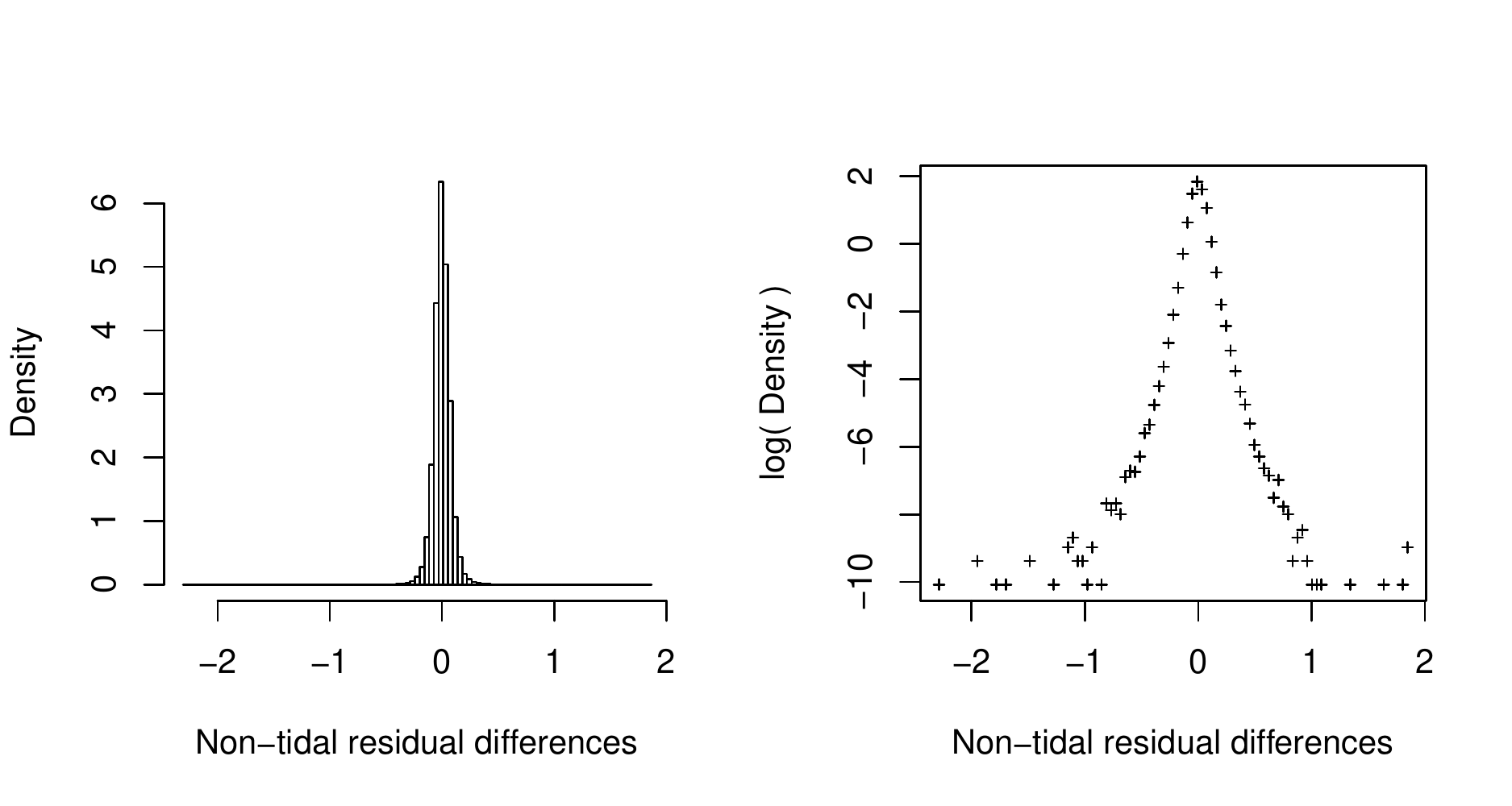}
\end{center}
\caption{Histogram of non-tidal residual differences at Heysham
(in a logarithmic scale at the left).}
\label{fig:10}
\end{figure}

\begin{figure}[t]
\begin{center}
\includegraphics[width=0.38\textwidth]{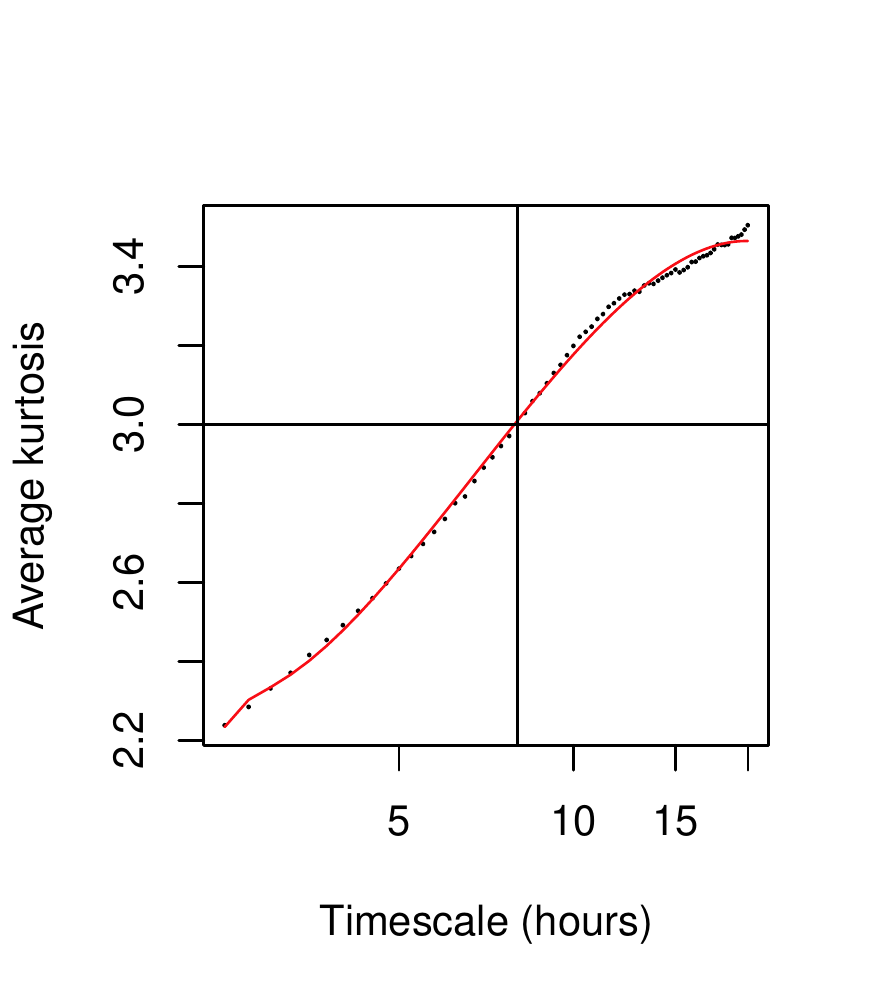}
\includegraphics[width=0.38\textwidth]{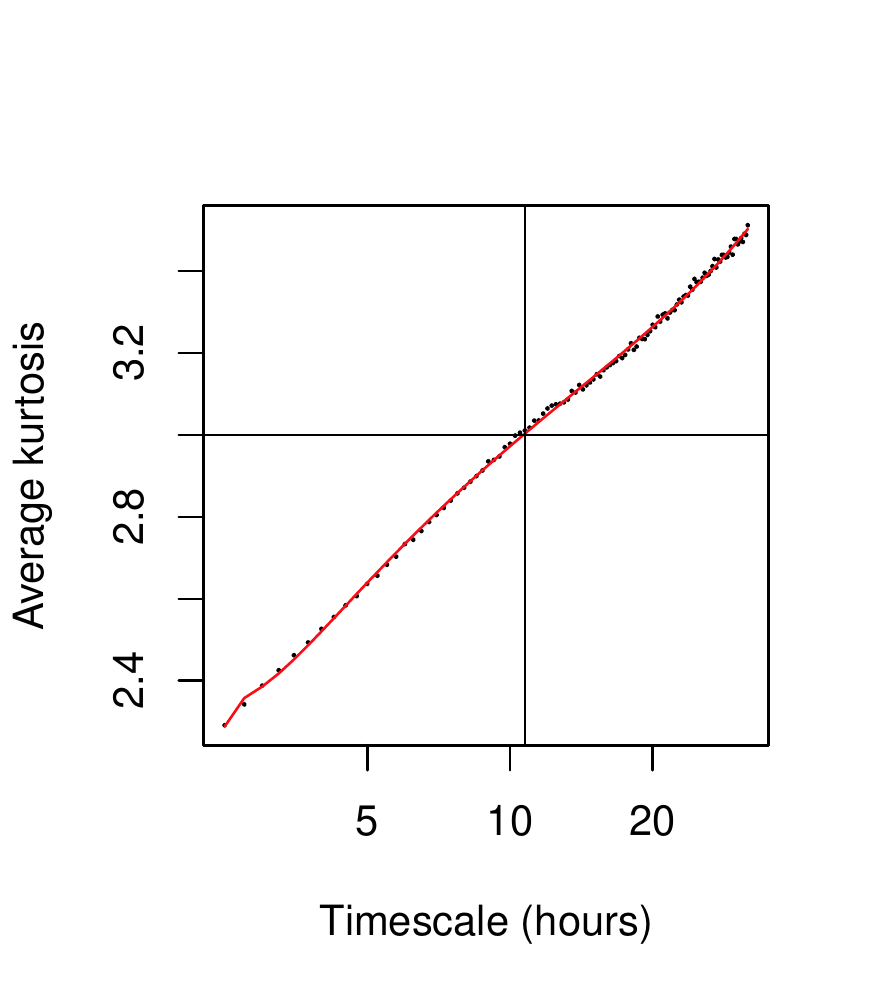}
\end{center}
\caption{As Figure \ref{fig:5} but for non-tidal residual differences instead of
residuals.}
\label{fig:11}
\end{figure}

\begin{figure}[t]
\begin{center}
\includegraphics[width=0.76\textwidth]{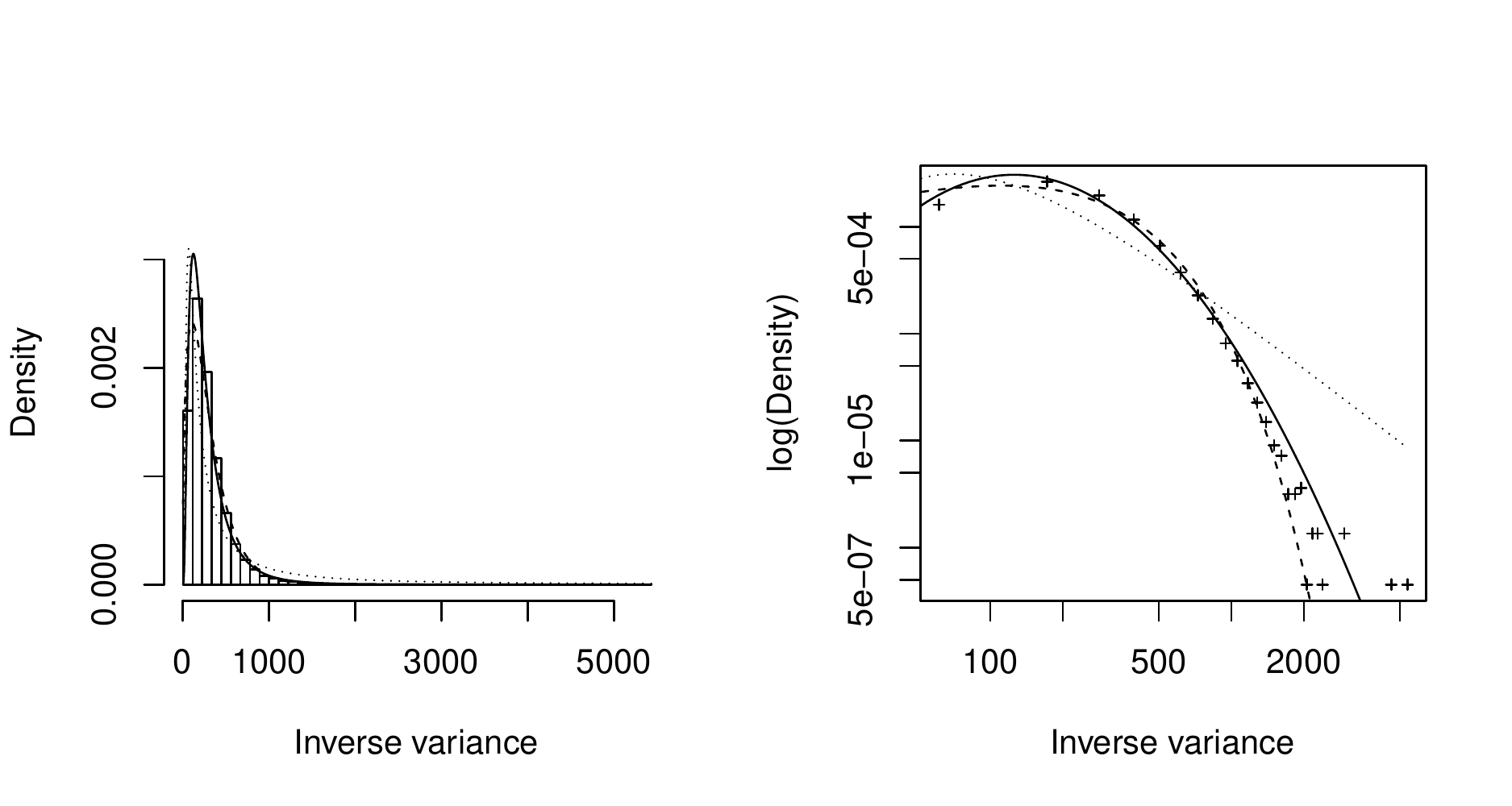}
\end{center}
\caption{Probability density $f(\beta)$ extracted form the non-tidal
residual differences $Y(t_{i+1}) - Y(t_i)$ at Dover and compared
to log-normal (solid line), $\chi^2$ (dashed line) and
inverse-$\chi^2$ (dotted line). Left:linear-linear scale, right:log-log scale.
All distributions have the same mean and variance as the experimental data.}
\label{fig:12}
\end{figure}

\begin{figure}[t]
\begin{center}
\includegraphics[width=0.76\textwidth]{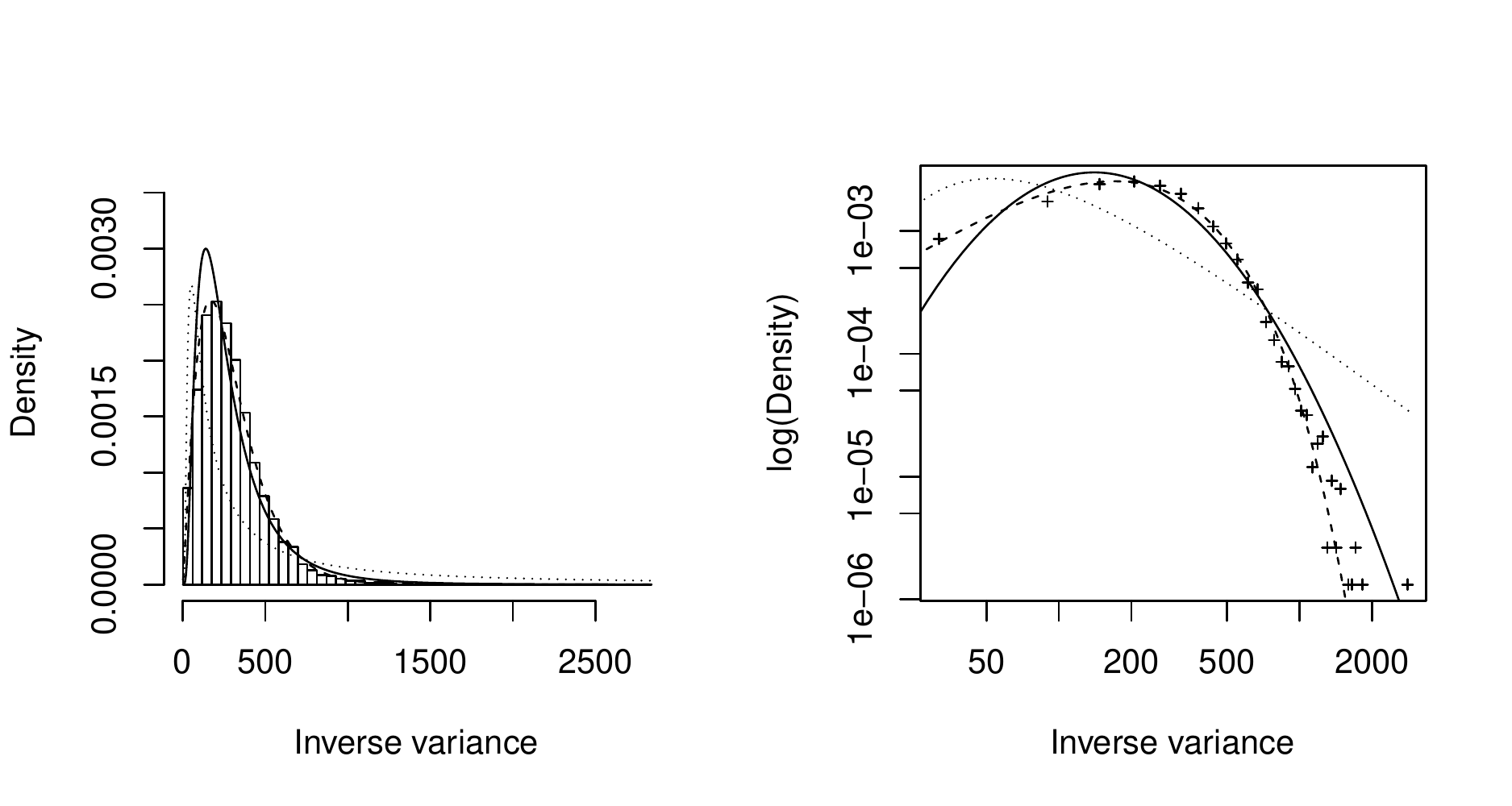}
\end{center}
\caption{As Figure \ref{fig:12} but for non-tidal residual differences at Heysham.}
\label{fig:13}
\end{figure}

\begin{table}[t]
\begin{center}
 \begin{tabular}{| l | r | r | r | r | r | r | c | c | c |}
 \hline
\multirow{2}{*}{Location} & \multirow{2}{*}{ $\tau$} & \multirow{2}{*}{$T$ } &
\multirow{2}{*}{$\tau/T$}& \multirow{2}{*}{$|\epsilon|$} & \multirow{2}{*}{ Skewness} &
\multirow{2}{*}{ Kurtosis}  & \multicolumn{3}{ c|}{ K-S Distance} \\
 \cline{8-10}
  & & & & & & & log-norm. & $\chi^2$ & Inv-$\chi^2$ \\
 \hline
 Dover     & $1$ & $8$ & $0.125$& $0.12$
& $0.10$ & $7.61$ & $0.033$ & $0.051$ & $0.212$  \\
 Heysham   & $0.75$ & $10.75$ & $0.070$& $0.72$
& $0.01$ & $16.31$ & $0.060$ & $0.020$ & $0.363$  \\
 Liverpool & $0.75$ & $11.75$ & $0.064$& $0.34$
& $-0.17$ & $12.81$ & $0.063$ & $0.018$ & $0.367$ \\
 Newlyn    & $0.25$ & $13.5$ & $0.018$& $0.25$
& $0.10$ & $10.91$  & $0.087$ & $0.038$ & $0.367$ \\
 Sheerness & $1.25$ & $35.25$& $0.035$& $0.09$
& $0.06$ & $6.41$ & $0.062$ & $0.019$ & $0.188$\\
 \hline
\end{tabular}
\end{center}
\caption{Relevant statistical parameters for applicability of the superstatistical
model to differences of observed non-tidal residuals at different locations.
The short and the long time scales ($\tau$ and $T$) are expressed in hours.
Parameters $\tau$, $T$ and $\epsilon$ are described in Section \ref{sec:tools}.}
\label{tb:3}
\end{table}

In the previous section we applied the superstatistical analysis
directly to the non-tidal residual but we encountered with the
problem that the skewness of the data disagrees with the underlying
hypothesis of the simplest superstatistical model, which is based on local Gaussian behavior.

Similar as in hydrodynamic turbulence, where one often considers velocity {\em differences} rather
than velocities itself as the relevant quantities, let us therefore look at
the differences of observed non-tidal residuals, $Y(t_{i+1}) - Y(t_i)$,
instead of the residuals $Y(t_{i})$ themselves. Figure \ref{fig:8}
displays the available data at Heysham. Figures \ref{fig:9} and
\ref{fig:10} display the histogram of differences at
Dover and Heysham respectively.

In Table \ref{tb:3} we show all relevant parameters for the
superstatistical analysis of non-tidal residual differences. We can observe
that the data is no longer skewed and the kurtosis is still bigger than $3$, as it
should be for the entire time series.
In all cases except Dover  the ratio $\tau/T$ is smaller that
in Table \ref{tb:2}, which implies that the time scale separation
is stronger in this case. On the other hand, the parameter
$|\epsilon|$ is in general bigger but it is still acceptable.
Cases like Heysham, where the parameter $|\epsilon|$ is relatively big,
are somewhat exceptional. In \cite{straeten} it was shown that the
presence of outliers in the data set causes $|\epsilon|$ to grow. The data set
considered here may contain measurement errors.
Due to the continuous nature of the sea-level variation, each observation in the data
set should be close to the previous one, but possible errors in the data are magnified
when considering the difference between sea levels. We believe that this 
could be a reason why the parameter $|\epsilon|$ is generally bigger than in the previous
case. But further analysis is needed to clarify why the Heysham data generate particularly
big $\epsilon$.

Our analysis summarized in Table \ref{tb:3} shows that $\chi^2$-superstatistics is best
suited to model the empirical distribution of sea-level changes, better than
inverse $\chi^2$-superstatistics or lognormal superstatistics,
see also Figures \ref{fig:12} and \ref{fig:13}. This means that on a long time
scale $q$-statistics is a good description of the data, and thus the corresponding
probability densities maximize $q$-entropies
\cite{tsallis1, tsallis-book}. Note that the only case
where $\chi^2$ superstatistics appears to be not the best model (Dover) is also the case with the
weakest time scale separation.

\section{Conclusion}

In this paper we have analyzed time series of measured sea level heights for different locations in the UK.
While there are minor differences for different locations, the overall picture is that
the stochastic component of sea-level changes is well-described by a superstatistical model
of $\chi^2$-type. This type of behavior is different from e.g. fully
developed hydrodynamic turbulence, which is better described by
lognormal superstatistics \cite{swinney}, or wind velocity fluctuations,
described by inverse $\chi^2$ superstatistics \cite{rapisarda,kantz}.
In future work one might aim to develop a generalized
statistical mechanics for the complex behavior of these types of observables in complex
environmental systems. A generalized statistical mechanics to describe the statistics
of sea levels would thus need to be based on $\chi^2$ superstatistics (effectively
leading to $q$-statistics), as shown in this paper by careful analysis of the available data.

\section*{Acknowledgements}

This research was supported by the EPSRC grant `Flood MEMORY'. The research
was also supported in part by the National Science Foundation
under Grant No. NSF PHY11-25915.


\begin{thebibliography}{99}
\bibitem{beck-cohen} C. Beck and E.G.D. Cohen,  Physica A {\bf 322}, 267 (2003)
\bibitem{swinney} C. Beck, E.G.D. Cohen, and H.L. Swinney, Phys. Rev. E {\bf 72}, 056133 (2005)
\bibitem{touchette} H. Touchette and C. Beck,
Phys. Rev. E {\bf 71}, 016131 (2005)
\bibitem{jizba} P. Jizba, H. Kleinert,
Phys. Rev. E {\bf 78}, 031122 (2008)
\bibitem{chavanis} P.-H. Chavanis,
Physica A {\bf 359}, 177 (2006)
\bibitem{frank} S.A. Frank and D.E. Smith, Entropy {\bf 12}, 289 (2010)
\bibitem{celia} C. Anteneodo and S.M. Duarte Queiros, J. Stat. Mech. P10023 (2009)
\bibitem{straeten} E. Van der Straeten and C. Beck, Phys. Rev. E {\bf 80}, 036108 (2009)
\bibitem{mark} C. Mark, C. Metzner, B. Fabry, arXiv:1405.1668
\bibitem{hanel} R. Hanel, S. Thurner, and M. Gell-Mann,
PNAS {\bf 108}, 6390 (2011)
\bibitem{tsallis1} C. Tsallis, J. Stat. Phys. {\bf 52}, 479 (1988)
\bibitem{souza} C. Tsallis and A.M.C. Souza,
Phys. Rev. E {\bf 67}, 026106 (2003)
\bibitem{briggs} K. Briggs, C. Beck,
Physica A {\bf 378}, 498 (2007)
\bibitem{prl} C. Beck,
Phys. Rev. Lett. {\bf 98}, 064502 (2007)
\bibitem{chen} L. Leon Chen, C. Beck,
Physica A {\bf 387}, 3162 (2008)
\bibitem{abul-magd} A.Y. Abul-Magd, G. Akemann, P. Vivo, J. Phys. A Math. Theor. {\bf 42}, 175207 (2009)
\bibitem{daniels} K.E. Daniels, C. Beck, and E. Bodenschatz,
Physica D {\bf 193}, 208 (2004)
\bibitem{cosmic} C. Beck,
Physica A {\bf 331}, 173 (2004)
\bibitem{rapisarda} S. Rizzo and A. Rapisarda,
AIP Conf. Proc. {\bf 742}, 176 (2004)
\bibitem{soby} D.N. Sobyanin, Phys. Rev. E {\bf 24}, 051128 (2011)
\bibitem{dixit} P.D. Dixit, arXiv:1210.3015
\bibitem{tsallis-book} C. Tsallis, Introduction to
Nonextensive Statistical Mechanics, Springer, 2009
\bibitem{kantz} M.S. Santhanam and H. Kantz, Phys. Rev. E {\bf 78},
051113 (2008)
\bibitem{yalcin} G.C. Yalcin, C. Beck. Physica A {\bf 392} 21 (2013)
\bibitem{porporato} A. Porporato, G. Vico, P.A. Fay, Geophys. Res. Lett. {\bf 33}, L15402 (2006)
\bibitem{tawn} J.A. Tawn. J. R. Stat. Soc. Series C, {\bf 41}, 1 (1992)

\bibitem{merrifield}
M.A. Merrifield, M. E. Maltrud, Geophys. Res. Lett., {\bf 38}, L21605 (2011)
\bibitem{calafat} F. M. Calafat, D. P.  Chambers, M.N. Tsimplis
J. Geophys. Res., {\bf 117}, C09022  (2012)
\bibitem{dangendorf1} S. Dangendorf, C. Mudersbach, T. Wahl, J. Jensen.
Ocean. Dyn., {\bf 63}, 209-224, (2013)
\bibitem{dangendorf2} S. Dangendorf, S. Müller-Navarra, J. Jensen, F. Schenk,
T. Wahl, and R. Weisse, J. Climate, {\bf 27} (2014)
\bibitem{horsburgh} K.J. Horsburgh, C. Wilson. J. Geophys. Res., {\bf 112},
C08003 (2007)
\bibitem{peltier}
W. R. Peltier, Ann. Rev. Earth. Planet. Sci., {\bf 32}, 111-149 (2004)
\bibitem{woppelmann}
G. W\"oppelmann, C. Letetrel, A. Santamaria, M-N. Bouin, X. Collilieux,
Z. Altamimi, S.D.P. Williams, B. Martin Miguez, Geophys. Res. Lett., {\bf 36},
L12607 (2009)
\bibitem{cabanes} C. Cabanes, A. Cazenave, C. Le Provost. Science. {\bf 294}, 5543 (2001)
\bibitem{haigh} I. Haigh, R. Nicholls, N. Wells. Cont. Shelf Res. {\bf 29}, 17 (2009)
\bibitem{pawlowicz} R. Pawlowicz, B. Beardsley, S. Lentz.
Comput. Geosci. {\bf 28}, 8 (2002)

\bibitem{batstone} C. Batstone, M. Lawless, J. Tawn, K. Horsburgh,
D. Blackman, A. McMillan, D. Worth, S. Laeger, T. Hunt.
Ocean Eng., {\bf 71} (2013)
\bibitem{eadie} W. T. Eadie, D. Drijard, F.E. James, M. Roos, B. Sadoulet.
Statistical Methods in Experimental Physics (1971)


\bibitem{BODC} British Oceanographic Data Centre. \url{http://www.bodc.ac.uk/}
\bibitem{hydro} C. Beck, Physica A 277, 115 (2000)
\end{thebibliography}

\end{document}